\documentclass[a4paper]{article}

\usepackage{lmodern}
\usepackage[T1]{fontenc}
\usepackage[utf8]{inputenc}
\usepackage{amssymb,amsmath}

\usepackage[hyphens]{url}

\usepackage{hyperref}
\hypersetup{unicode=true,
            pdfborder={0 0 0},
            breaklinks=true}
\usepackage{longtable,booktabs}
\IfFileExists{parskip.sty}{%
\usepackage{parskip}
}{%
\setlength{\parindent}{0pt}
\setlength{\parskip}{6pt plus 2pt minus 1pt}
}
\usepackage{natbib}
\usepackage{hyperref}
\usepackage{nameref}
\usepackage{graphicx}  %
\usepackage{multirow}
\usepackage{tikz}
\usetikzlibrary{shapes,arrows,positioning,automata}
\newcommand{\snapper}[0]{\textsc{SnapperGPS}}
\definecolor{sea}{RGB}{211,219,219}
\usepackage{overpic}
\usepackage{siunitx}    %
\DeclareSIUnit\year{yr}
\DeclareSIUnit\day{d}
\DeclareSIUnit\bit{bit}
\DeclareSIUnit\byte{B}

\usepackage{subcaption}         %
\usepackage[export]{adjustbox}  %

\title{\snapper{}: Open Hardware for Energy-Efficient, Low-Cost Wildlife Location Tracking with Snapshot GNSS}
\usepackage{authblk}
\author[1]{Jonas Beuchert} 
\author[1]{Amanda Matthes} 
\author[1]{Alex Rogers}
\affil[1]{University of Oxford, UK}

\begin{document}
\maketitle

\begin{abstract}
Location tracking with global navigation satellite systems (GNSS), such as GPS, is used in many applications, including the tracking of wild animals for research. Snapshot GNSS is a technique that only requires milliseconds of satellite signals to infer the position of a receiver. This is ideal for low-power applications such as animal tracking. However, there are few existing snapshot systems, none of which is open source. 

To address this, we developed \snapper{}, a fully open-source, low-cost, and low-power location tracking system designed for wildlife tracking. \snapper{} comprises three parts, all of which are open-source: (i) a small, low-cost, and low-power receiver; (ii) a web application to configure the receiver via USB; and (iii) a cloud-based platform for processing recorded data. This paper presents the hardware side of this project.

The total component cost of the receiver is under \$30, making it feasible for field work with restricted budgets and low recovery rates. The receiver records very short and low-resolution samples resulting in particularly low power consumption, outperforming existing systems. It can run for more than a year on a \SI{40}{\milli \ampere \hour} battery. 

We evaluated \snapper{} in controlled static and dynamic tests in a semi-urban environment where it achieved median errors of \SI{12}{\metre}. Additionally, \snapper{} has already been deployed for two wildlife tracking studies on sea turtles and sea birds.

\end{abstract}

\begin{longtable}[]{@{}l@{}}
\begin{minipage}[t]{0.97\columnwidth}\raggedright\strut

\section{Metadata Overview}\label{sec:metadata_overview}%

Main design files: \url{https://github.com/SnapperGPS/snappergps-pcb}

Target group: biologists tracking animal movement

Skills required: PCB manufacturing and assembly (can be outsourced) - advanced; 

Replication: This hardware has been replicated by every author.

See section ``Build Details'' for more detail.

\section{Keywords}\label{sec:keywords}%

{\noindent conservation technology; wildlife tracking; satellite navigation; low power; low cost; open source;}

\strut\end{minipage}\tabularnewline
\bottomrule
\end{longtable}

\section{(1) Overview} \label{sec:overview}

\subsection{Introduction}\label{sec:overview:introduction}

Animal location trackers most commonly use global navigation satellite systems (GNSS) such as the Global Positioning System (GPS). These systems use constellations of satellites which continuously broadcast radio signals containing precise transmission timestamps and their ephemeris data describing their orbits. A GNSS tracking device on Earth captures these signals and infers its location from this information. 
A traditional GNSS receiver usually requires more than one minute of data for a first fix (for which it needs to decode both, ephemerides and timestamps), and at least several seconds for subsequent fixes (if recently decoded ephemerides are available) \citep{diggelen2009}. However, powering the radio to capture these signals and then the processor to calculate the location is energy expensive, resulting in the need for large batteries with high capacity. This often makes traditional GNSS tags impractical for tracking small animals over long deployment periods \citep{mcmahon2017}.

Assisted GNSS (A-GNSS) receivers address this issue by obtaining some of the satellite data in another way, which allows them to reduce their on-time, thereby saving energy. This is done either by pre-loading ephemerides before the deployment or by regularly downloading them via another connection (e.g. cellular). However, pre-loading data is only possible for short deployments and an additional download link requires more expensive hardware and is energy intensive.

Snapshot GNSS is another alternative GNSS concept, which, by design, has significantly lower energy needs, resulting in small, light-weight, energy-efficient, and low-cost receivers. Instead of capturing seconds or even minutes of the satellite signals for a fix, a snapshot receiver records just a few milliseconds. This reduces its power consumption by several orders of magnitude compared to a traditional GNSS approach. Additionally, a snapshot GNSS device does not need to calculate its position on-board, saving even more energy and lowering the requirements for its compute hardware. Instead, it can just locally store the raw signal snapshots until the deployment ends. Afterwards, the data processing can be off-loaded to the cloud.

\begin{table}[t]
\caption{Existing snapshot GNSS systems.}
\begin{center}
\label{tab:snapshot}
\resizebox{\columnwidth}{!}{%
\begin{tabular}{l c c c c}
\textbf{}           & \textbf{CO-GPS}       & \textbf{Baseband Technologies}    & \textbf{ETH Z\"{u}rich}   &  \textbf{ATS G10 Ultralite GPS} \\
\hline
\tiny &&&& \\
\textbf{Reference}  & \cite{liu2012}        & \cite{baseband2020}               & \cite{eichelberger2019}   & \cite{australia2020} \\
\tiny &&&& \\
\textbf{Memory}     & 8 MBit                & 4 GB                              & 2 GB                      & 4 GB \\
                    & $\leq$ 1,000 snapshots  & $\leq$ 2,000,000 snapshots          & 65,600 snapshots            & $\leq$ 244,000 snapshots    \\
&&&& \\
\textbf{Maximum}    & 1.5~years             & 18~days--1~year                   & 683~days                  & 80~min \\
\textbf{deployment} & (2~AA~batteries,      & (10~mAh)                   & (coin cell,               & (19~mAh, 1~fix/s) \\
\textbf{duration}   &  1~fix/s)             & weeks (coin cell)                 & 235~mAh,             & --759~days \\
                    &                       & years (phone battery)             & 4~fixes/h)                & (200~mAh, 1~fix/h) \\
&&&& \\
\textbf{Snapshot}   & 5 chunks of 2 ms      & 2--20 ms                          & 1--30 ms& ? \\
\textbf{duration}   & & & \\
&&&& \\
\textbf{Quantisation}& 2x2 bit              & ?                                 & 2 bit                     & ? \\
&&&& \\
\textbf{Sampling}   & 16.368 MHz            & ?                                 & 16.368 MHz                & ? \\
\textbf{frequency}  & & & & \\
&&&& \\
\textbf{Accuracy}   & $<$ 35 m$^\mathrm{a}$ & median $<$ 9 m                    & $<$ 25 m$^\mathrm{b}$     & mean 15.5 m$^\mathrm{c}$ \\
&&&& \\
\textbf{Weight}     & ?                     & ?                                 & 1.3~g                     & 11~g \\
&&&& \\
\textbf{Size [mm]}       & 70$\times$52       & 22$\times$27                   & 23$\times$14           & 32$\times$23$\times$12 \\
&&&& \\  %
\textbf{Available (2022)}& no               & yes$^\mathrm{d}$                  & no                        & no \\
&&&& \\
\textbf{Price}      & N/A                   & 189 USD                           & N/A                       & N/A \\
&&&& \\
\textbf{Open source}& no                    & no                                & no                        & no \\
\hline
\multicolumn{5}{l}{$^{\mathrm{a}}$Data from commercial GPS front-end, not snapshot receiver.} \\
\multicolumn{5}{l}{$^{\mathrm{a}}$Rooftop with good sky visibility.} \\
\multicolumn{5}{l}{$^{\mathrm{c}}$Evaluation by \cite{mcmahon2017}.} \\
\multicolumn{5}{l}{$^{\mathrm{d}}$Evaluation board.}
\end{tabular}
}
\end{center}
\end{table}

Table~\ref{tab:snapshot} details pure snapshot GNSS systems that have been developed in the past. Only \textsc{Baseband Technologies} currently offers its solution for purchase \citep{baseband2020}. It is available as a proprietary development board and, therefore, not readily available for deployments. Only the ATS G10 module has been evaluated in a realistic scenario, but was found to be unreliable by \cite{mcmahon2017} due to battery and software failures \citep{australia2020}. Furthermore, all existing solutions also use high sample rates and/or long snapshots at multi-bit resolution. This improves the snapshot quality but also requires complex and expensive hardware. For example, the \textsc{ETHZ} receiver records two-bit signals sampled at \SI{16}{\mega \hertz}, which limits their microcontroller choice to one with a parallel input capture interface (PARC) \citep{eichelberger2019}. The \textsc{CO-GPS} receiver uses additional circuitry to convert its two-bit GPS input stream at \SI{16}{\mega \hertz} into a 16-bit parallel signal at \SI{2}{\mega \hertz}, which a microcontroller then captures using direct memory access (DMA) \citep{liu2012}.

Moreover, the hardware is only part of a full snapshot GNSS solution. A complete system also needs a signal processing chain to calculate positions from the raw snapshot data. However, this software is not openly available for any of these systems, which renders users dependent on the technology provider over the whole lifespan of their devices.

To address these issues, we developed \snapper{}, a completely open-source, small, low-cost, low-power snapshot GNSS system designed for tracking wildlife. The algorithmic details of the \snapper{} cloud-processing chain have already been described \citep{beuchert2021b}. This paper presents the accompanying hardware. The \snapper{} printed circuit board (PCB) measures \SI{32.0}{\milli \metre} by \SI{27.3}{\milli \metre} and weighs \SI{3}{\gram}. The total weight, including an antenna and a \SI{40}{\milli\ampere\hour} lithium-ion polymer battery, is \SI{9}{\gram}. With such a battery \snapper{} can run for more than a year. \snapper{} works with short \SI{12}{\milli \second} snapshots sampled at only \SI{4}{\mega \hertz} with \SI{1}{\bit} amplitude quantisation. This unmatched low resolution allows us to use low-cost, off-the-shelf components. The total component cost adds up to \$21 for each device if ordered in a batch of 100, excluding battery and antenna. \snapper{} tags are used with an open-source web application. The app serves as a tool for configuring the tag and later for uploading the recorded snapshots to the cloud for processing. We evaluated \snapper{} in controlled deployments, both moving and stationary. With good sky visibility, \snapper{} achieves a median error of \SI{10}{\metre}. Modified versions of \snapper{} have additionally been deployed on free-ranging sea turtles and sea birds, demonstrating its usefulness for wildlife tracking.

The rest of the paper is structured as follows:
Section \nameref{sec:overview} continues with the \nameref{sec:overview:implementation} of \snapper{}. It includes sub-sections on the \nameref{sec:overview:implementation:hardware}, the \nameref{sec:overview:implementation:firmware} and the \nameref{sec:overview:implementation:web_application}. Section \nameref{sec:quality_control} covers how to safely operate a \snapper{} receiver and how to test that the system reliably provides accurate location estimates over long periods of time. Examples of use cases are presented in \nameref{sec:application} and \nameref{sec:build_details} provides the instructions to replicate \snapper{}. Section \nameref{sec:discussion} includes a conclusion and an outlook on future work.

\subsection{Overall implementation and design}\label{sec:overview:implementation}

\snapper{} consists of three components: (i) a purpose-built energy-efficient low-cost receiver, (ii) a web application for configuration of the receiver, and (iii) a cloud-based data processing platform.

\subsubsection{Electronics}\label{sec:overview:implementation:hardware}

\begin{figure}[tb]
\centerline{\begin{overpic}[width=0.5\columnwidth]{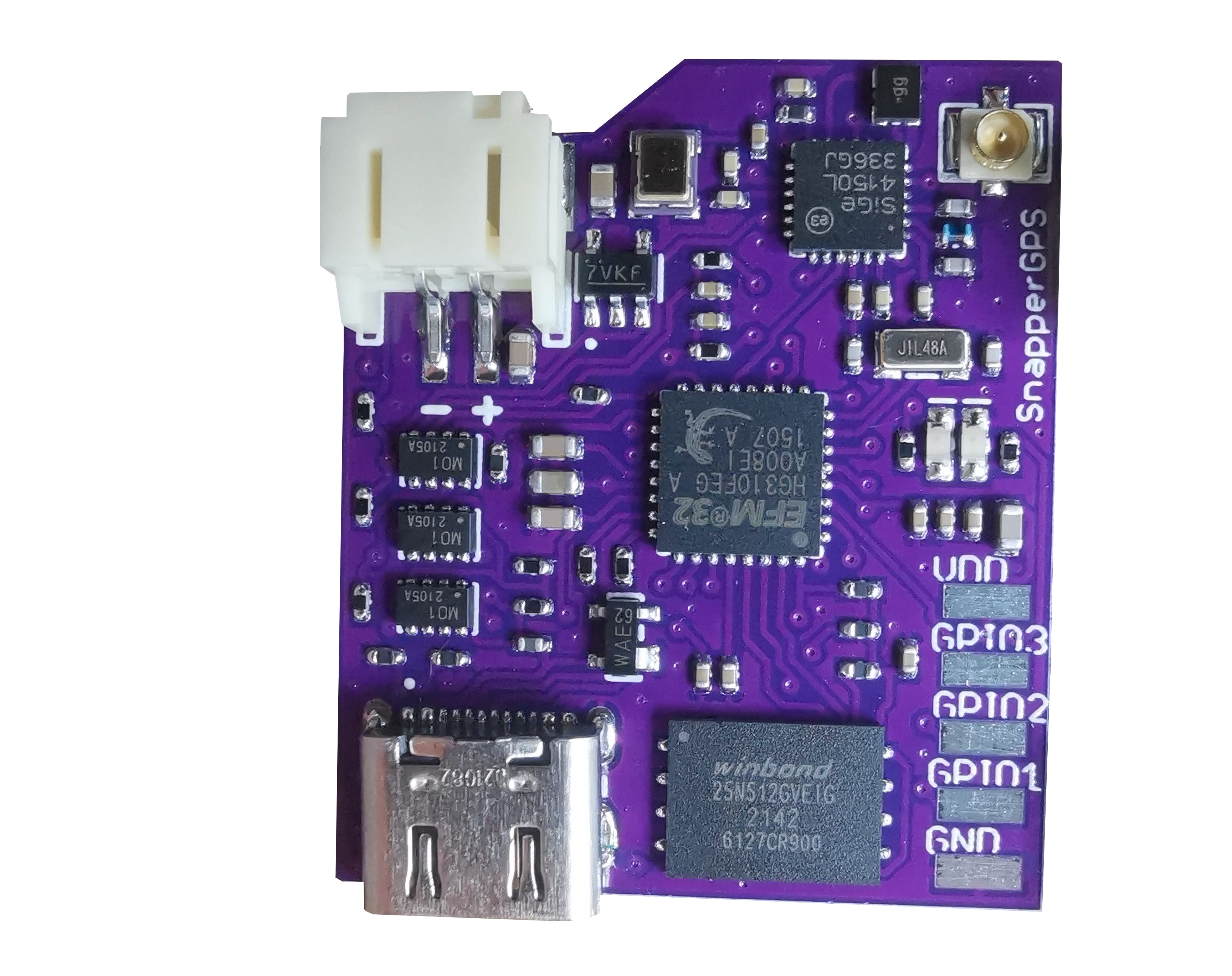}
\put(40,2){\vector(-1,0){12}}
\put(48,0){\SI{27.3}{\milli\metre}}
\put(72,2){\vector(1,0){12}}
\put(92,19){\vector(0,-1){12}}
\put(90,28){\rotatebox{90}{\SI{32.0}{\milli\metre}}}
\put(92,61){\vector(0,1){12}}
\end{overpic}}
\caption{Assembled \snapper{} receiver.}
\label{fig:pcb}
\end{figure}

\begin{figure}[tb]

\centerline{\begin{overpic}[width=0.8\columnwidth]{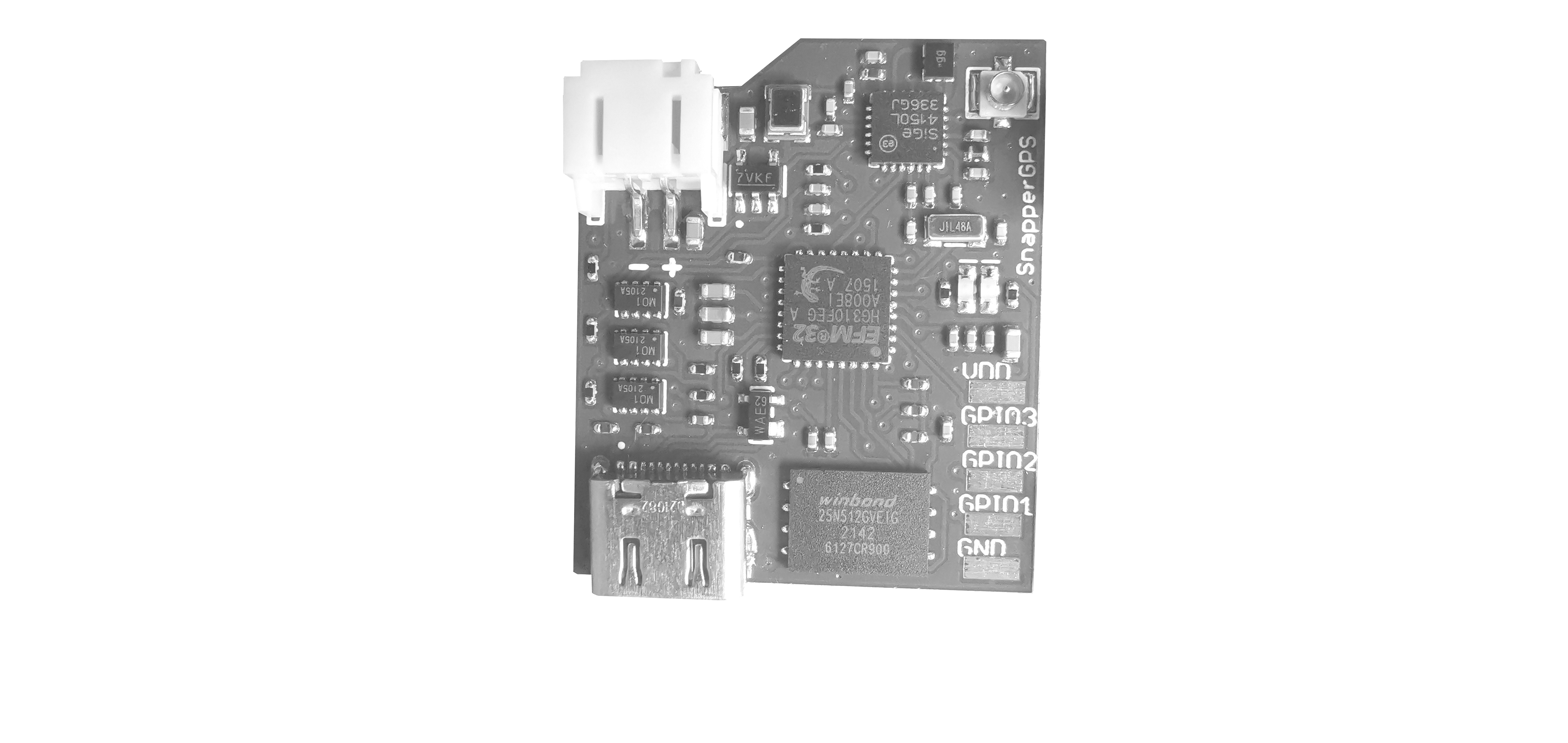}

\put(-8, 40){\SI{16.368}{\mega \hertz} oscillator}
\put(28, 41){\color{violet}\vector(1,0){21.5}}

\put(-10, 35){JST battery connector}
\put(28, 36){\color{violet}\vector(1,0){9.5}}

\put(17, 30){MCU}
\put(28, 31){\color{violet}\vector(1,0){22}}

\put(-4, 25){power management}
\put(28, 26){\color{violet}\vector(1,0){13}}

\put(-2, 10){USB-C connector}
\put(28, 11){\color{violet}\vector(1,0){11}}

\put(75, 11){flash memory}
\put(73, 12){\color{violet}\vector(-1,0){15.5}}

\put(75, 17){exposed GPIO pins}
\put(73, 18){\color{violet}\vector(-1,0){9}}

\put(75, 27){LEDs}
\put(73, 28){\color{violet}\vector(-1,0){10}}

\put(75, 31.5){\SI{32.768}{\kilo \hertz} oscillator}
\put(73, 32.5){\color{violet}\vector(-1,0){11}}

\put(75, 36){radio}
\put(73, 37){\color{violet}\vector(-1,0){13}}

\put(75, 39.5){U.FL antenna connector}
\put(73, 40.5){\color{violet}\vector(-1,0){8.5}}

\put(75, 43){filter}
\put(73, 44){\color{violet}\vector(-1,0){13}}

\end{overpic}}

\tikzstyle{block} = [
    draw,
    fill=gray!40,  %
    rectangle, 
    minimum height=3em,
    minimum width=6em
]

\resizebox{\columnwidth}{!}{%

\begin{tikzpicture}[auto, node distance=8em, >=latex']%

    \node [block] (filter) {filter};
    \node [block, right of=filter] (radio) {radio};
    \node [block, right of=radio] (mcu) {MCU};
    \node [block, right of=mcu] (usb) {USB};
    \node [block, above of=filter, node distance=6em] (antenna) {antenna};
    \node [block, right of=antenna] (leds) {LEDs};
    \node [block, right of=leds, align=center] (power) {power\\management};
    \node [block, right of=power] (battery) {battery};
    \node [block, below of=radio, node distance=6em, align=center] (tcxo) {\SI{16.368}{\mega \hertz}\\oscillator};
    \node [block, right of=tcxo, align=center] (oscillator) {\SI{32.768}{\kilo \hertz}\\oscillator};
    \node [block, right of=oscillator] (memory) {memory};

    \draw (filter) -- (radio);
    \draw (radio) -- (mcu);
    \draw (mcu) -- (usb);
    \draw (filter) -- (antenna);
    \draw (mcu) -- (leds);
    \draw (mcu) -- (power);
    \draw (usb) -- (power);
    \draw (power) -- (battery);
    \draw (radio) -- (tcxo);
    \draw (mcu) -- (oscillator);
    \draw (mcu) -- (memory);

\end{tikzpicture}

}%
 
\caption{The main electronic components of a \snapper{} receiver.}
\label{fig:electronics_block_diagram}
\end{figure}

We place all electronic components on a single side of a two-layered PCB to enable low-cost assembly, see Figure~\ref{fig:pcb}.
The block diagram in Figure~\ref{fig:electronics_block_diagram} shows all core components, which we describe in the following.

The antenna captures GNSS signals in the GPS L1 band, which has a centre frequency of \SI{1.57542}{\giga \hertz}.
There are more GNSS signal bands, but the cheapest ICs work with the L1 band, the oldest civilian band, and we can receive the modernised GPS L1C, the Galileo E1, the BeiDou B1C, and the potential future GLONASS L1OCM signal in this band, too.
This allows us to make use of multiple satellite systems and, therefore, to increase localisation reliability by using more satellites.

Sampling and storing satellite signals at a rate of multiple gigahertz is not possible with a low-cost receiver.
Therefore, we use the heterodyne method: by mixing the received signal with an unmodulated signal from a local oscillator, we shift the original signal down to a lower, so-called intermediate frequency.
For this, we choose an integrated circuit that is designed for low-power applications, is available at very low costs (<\$1), and is designed for the GPS L1 band: the SE4150L-R from \textsc{Skyworks Solutions Inc.}
This \textit{radio} includes an analogue-to-digital converter (ADC) and a voltage-controlled oscillator (VCO), which needs a reference frequency of \SI{16.368}{\mega \hertz} from an external temperature-compensated crystal oscillator (TCXO).

To sample the radio's output signal, we connect it to a low-power microcontroller unit (MCU); a 32-bit \textsc{Silicon Labs Happy Gecko} with an \textsc{Arm Cortex-M0+} core.
To reduce the overall number of components, we operate the MCU on the same clock as the radio when acquiring snapshots.
Otherwise, we fall back to an internal \SI{14}{\mega \hertz} RC oscillator.

The MCU stores the sampled signals on an external flash memory chip.
Its size is the main limitation for the number of fixes.
We use a \SI{512}{\mega \bit} flash, which stores up to about 11,000~snapshots, \SI{6}{K\byte} each.

To configure the board or download data, it must be connected to a host device.
For this, we add a USB-C connector and use the internal USB hardware of the MCU.

The PCB also has two LEDs
to provide information on the state of the device when it is not connected via USB.

The board draws its power from either USB or a rechargeable lithium-ion polymer (LiPo) battery.
The latter's high power density allows to reduce weight
It provides \SIrange[range-phrase=--]{3.0}{4.2}{\volt}, while the USB power line is at \SI{5}{\volt}.  %
Therefore, we need a power regulator and a MOSFET-based switch to provide a stable \SI{3.3}{\volt} supply voltage.
There are also MOSFETs to switch off the radio and the flash to minimise quiescent currents.
Finally, we add a circuit to measure the battery voltage.

We expose three general-purpose input/output pins (GPIOs) of the MCU at an edge of the PCB to allow for the connection of external modules to extend the functionality of the device.
We also expose the reset pin and the Serial Wire Debug (SWD) pins.

\subsubsection{Firmware}\label{sec:overview:implementation:firmware}

\begin{figure}[t]

\tikzstyle{block} = [
    draw,
    fill=gray!40,  %
    circle,
    minimum width=7em
]
\tikzstyle{button} = [
    draw,
    fill={rgb:red,94;green,37;blue,144},  %
    text=white,
    rounded corners, 
    minimum height=1em,
    minimum width=5em
]

\centerline{

\begin{tikzpicture}[auto, node distance=11em, >=latex', every state/.style={thick, fill=gray!40, minimum width=7em},]%

    \node [state, align=center] (connected-unconfigured) {connected,\\unconfigured};
    \node [state, right of=connected-unconfigured, node distance=20em] (shutdown) {shutdown};
    \node [state, below of=connected-unconfigured, node distance=18em, align=center] (connected-configured) {connected,\\configured};
    \node [state, below of=shutdown, node distance=14em] (capture) {capture};
    \node [state, below of=capture, node distance=9em] (sleep) {sleep};

    \draw [->] (connected-configured) edge[above] node [below, xshift=-1.2em, yshift=-1.4em] {unplug} (capture);
    \draw [->] (connected-configured) edge[below] (sleep);
    \draw [->] (sleep) edge[right] node [above right, xshift=-0.6em, yshift=1em] {plug in} (connected-unconfigured);
    \draw [->] (capture) edge[right] node [right, align=left] {end time reached /\\memory full} (shutdown);
    \draw [->] (connected-unconfigured) edge[below, bend right] node [above] {unplug} (shutdown);
    \draw [<-] (connected-unconfigured) edge[above, bend left] node [below] {plug in} (shutdown);
    \draw [->] (capture) edge[left, bend left] (sleep);
    \draw [<-] (capture) edge[right, bend right] node [right, align=left, xshift=4.5em] {counter\\interrupt} (sleep);
    \draw [->] (connected-unconfigured) edge[left, bend left] node [button, above] {\footnotesize configure} (connected-configured);
    \draw [<-] (connected-unconfigured) edge[right, bend right] node [button, below] {\footnotesize shutdown} (connected-configured);

\end{tikzpicture}

}

\caption{State machine of of a \snapper{} receiver.}
\label{fig:firmware_state_machine}
\end{figure}
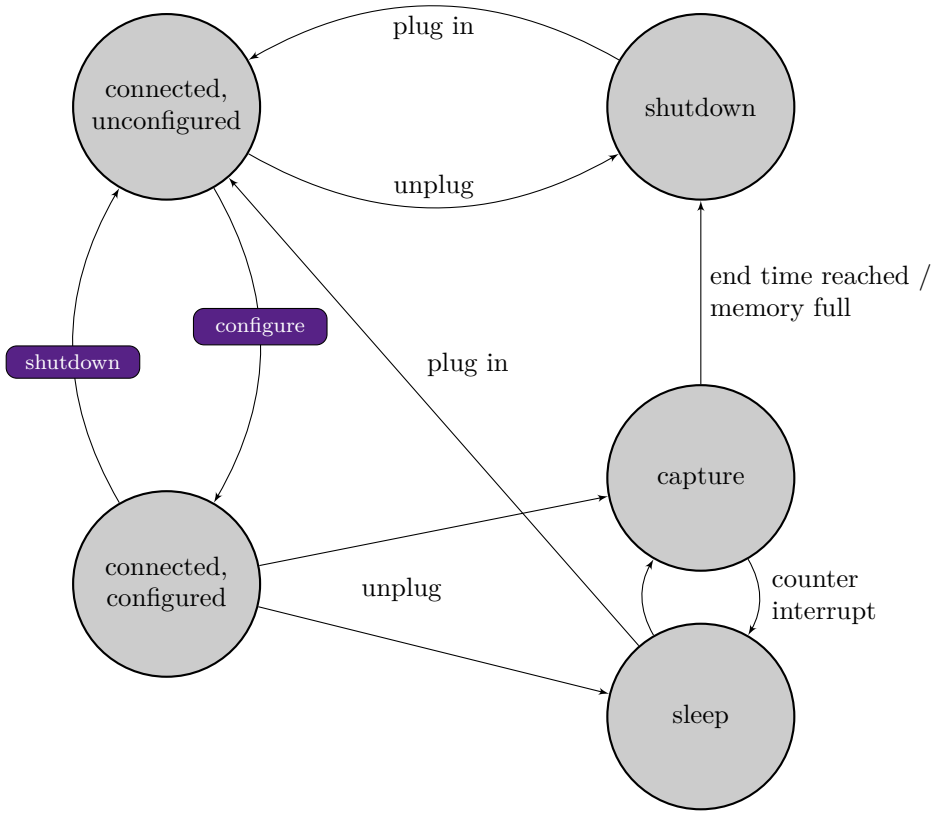

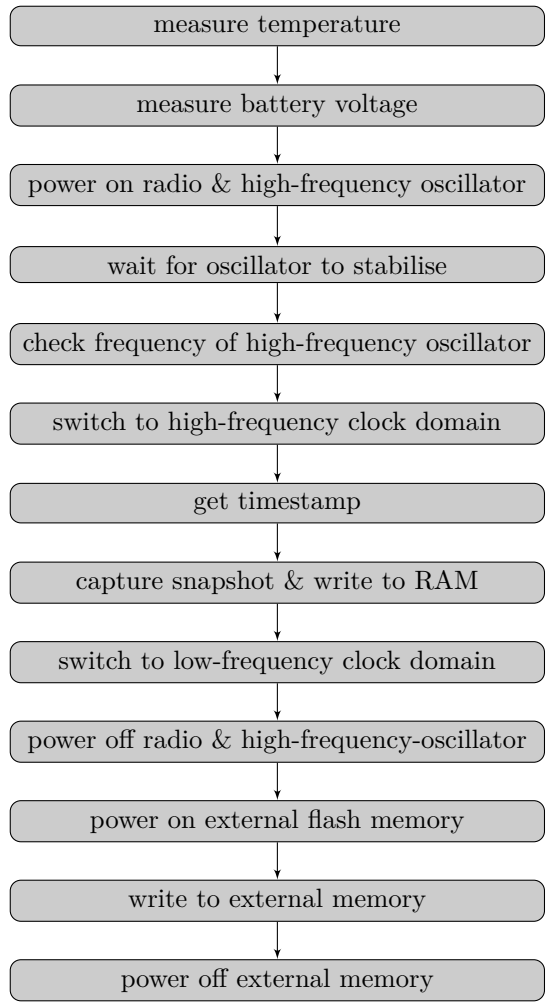
\begin{figure}[t]

\tikzstyle{block} = [
    draw,
    fill=gray!40,  %
    rounded corners,
    minimum width=20em
]

\centerline{

\begin{tikzpicture}[auto, node distance=3em, >=latex']%

    \node [block] (temp) {measure temperature};
    \node [block, below of=temp] (vbat) {measure battery voltage};
    \node [block, below of=vbat] (on) {power on radio \& high-frequency oscillator};
    \node [block, below of=on] (stabilise) {wait for oscillator to stabilise};
    \node [block, below of=stabilise] (check) {check frequency of high-frequency oscillator};
    \node [block, below of=check] (switch-high) {switch to high-frequency clock domain};
    \node [block, below of=switch-high] (time) {get timestamp};
    \node [block, below of=time] (snapshot) {capture snapshot \& write to RAM};
    \node [block, below of=snapshot] (switch-low) {switch to low-frequency clock domain};
    \node [block, below of=switch-low] (off) {power off radio \& high-frequency-oscillator};
    \node [block, below of=off] (flash-on) {power on external flash memory};
    \node [block, below of=flash-on] (write) {write to external memory};
    \node [block, below of=write] (flash-off) {power off external memory};

    \draw [->] (temp) -- (vbat);
    \draw [->] (vbat) -- (on);
    \draw [->] (on) -- (stabilise);
    \draw [->] (stabilise) -- (check);
    \draw [->] (check) -- (switch-high);
    \draw [->] (switch-high) -- (time);
    \draw [->] (time) -- (snapshot);
    \draw [->] (snapshot) -- (switch-low);
    \draw [->] (switch-low) -- (off);
    \draw [->] (off) -- (flash-on);
    \draw [->] (flash-on) -- (write);
    \draw [->] (write) -- (flash-off);

\end{tikzpicture}

}

\caption{Flow chart for capturing a snapshot.}
\label{fig:snapshot_capture_flow_chart}
\end{figure}

The
\textsc{Happy Gecko} MCU runs custom firmware written in the C programming language.
The state machine in Figure~\ref{fig:firmware_state_machine} visualises its core functionalities.
To minimise the overall energy consumption, we carefully select for each state one of the five energy modes (EMs) that the \textsc{Happy Gecko} offers
\citep{siliconlabs2015}.
Almost all the time during a deployment, the MCU sleeps in stop mode (EM3), which consumes only \SI{0.5}{\micro \ampere}.
A real-time counter (RTC) triggers wake-ups in regular intervals to record a satellite signal snapshot, respectively, and the MCU switches to run mode (EM0) for this.
The flow chart in Figure~\ref{fig:snapshot_capture_flow_chart} shows the steps that it completes in this mode before it goes to sleep again.

The radio outputs satellite signal samples at \SI{16.368}{\mega \hertz} with two-bit amplitude quantisation.
However, we sample the signal at a low \SI{4.092}{\mega \hertz} rate, a quarter of the clock frequency.
Additionally, we use only one bit for the amplitude quantisation.
Both together reduce the amount of required memory by a factor of eight.
However, the main advantage is that we can capture the one-bit receiver output using the existing device USART configured as an SPI master.
Our choice gives us 32~clock cycles to acquire an eight-bit sample and write it to RAM.
The small RAM size of our low-cost MCU limits the snapshot length to \SI{6}{K \byte}, i.e., \SI{12}{\milli \second}.
Afterwards, we
write the snapshot together with a timestamp and measurements of the battery voltage and the temperature via SPI to the external flash memory.
Subsequently, the MCU goes to sleep in EM3 again or switches to shutoff mode (EM4) if all snapshots have been captured and the deployment is complete.
In shutoff mode, the RTC is also powered down and the MCU consumes just \SI{20}{\nano \ampere}.

To preserve energy, we switch off the high-frequency \SI{16.368}{\mega \hertz} oscillator during sleeping because we only need a high frequency clock while capturing a snapshot and the low-frequency \SI{32.768}{\kilo \hertz} oscillator is sufficient to keep track of time between consecutive snapshots and the RTC running.

We configure a GPIO with a pull-up resistor to be pulled low when a USB connector is plugged in.
Then, the MCU leaves EM3 or EM4, stops the RTC and, therefore, the recording of further snapshots, and switches to EM0.
A \snapper{} receiver provides USB bulk endpoints in both directions to facilitate fast transfers of larger amounts of data.
These enable it to be configured as a WebUSB device, for example, together with the \snapper{} web application, which is linked in the USB descriptor.
The documentation in the snappergps-firmware repository defines the messages to communicate with a \snapper{} receiver via USB.
The messages can be used to get status information from the receiver,
 configure a receiver for a deployment, 
and transfer snapshots and timestamps to a host computer.
We also use USB to send a firmware update via the MCU to the external flash memory and to request a cyclic redundancy check (CRC) afterwards to detect transmission errors.
Next, a custom bootloader function that is held in SRAM copies the firmware update from the external flash memory to the flash memory of the MCU before the MCU reboots to complete the update.

\subsubsection{Cloud-processing back-end}

\begin{figure}[tb]
\tikzstyle{block} = [
    inner sep=0.4em,
    node distance=7.7em
]
\tikzstyle{close-block} = [
    inner sep=0.4em,
    node distance=4em
]
\tikzstyle{txt} = [
    node distance=3em,
    align=center,
    font=\footnotesize
]
\tikzstyle{arrow-txt} = [
    font=\footnotesize
]
\pgfdeclarelayer{background}
\pgfdeclarelayer{foreground}
\pgfsetlayers{background,main,foreground}
\begin{tikzpicture}[>=latex']
\node[block] (receiver)
    {\includegraphics[width=1.7em]{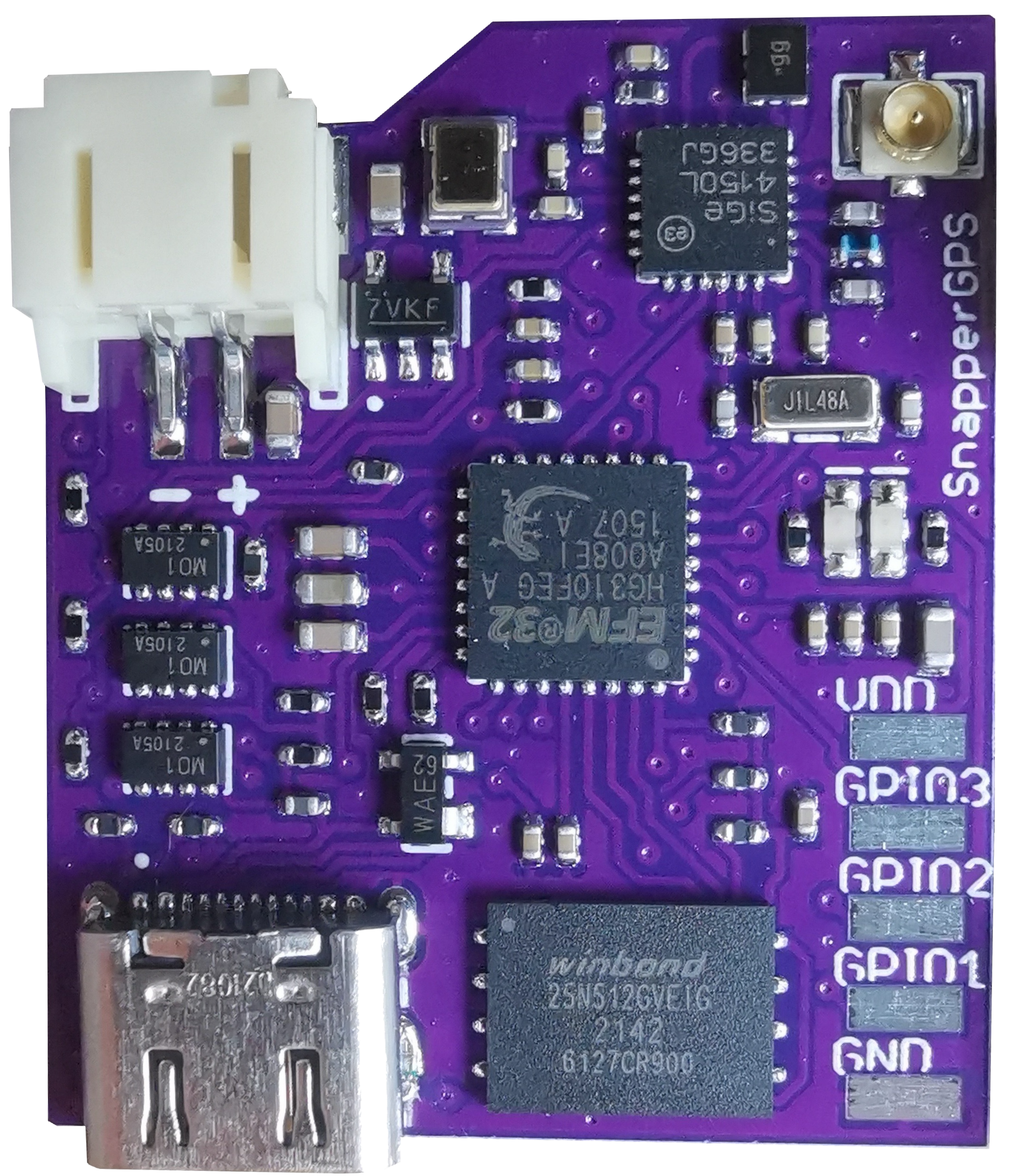}};
\node[block, right of=receiver] (app)
    {\includegraphics[width=2em]{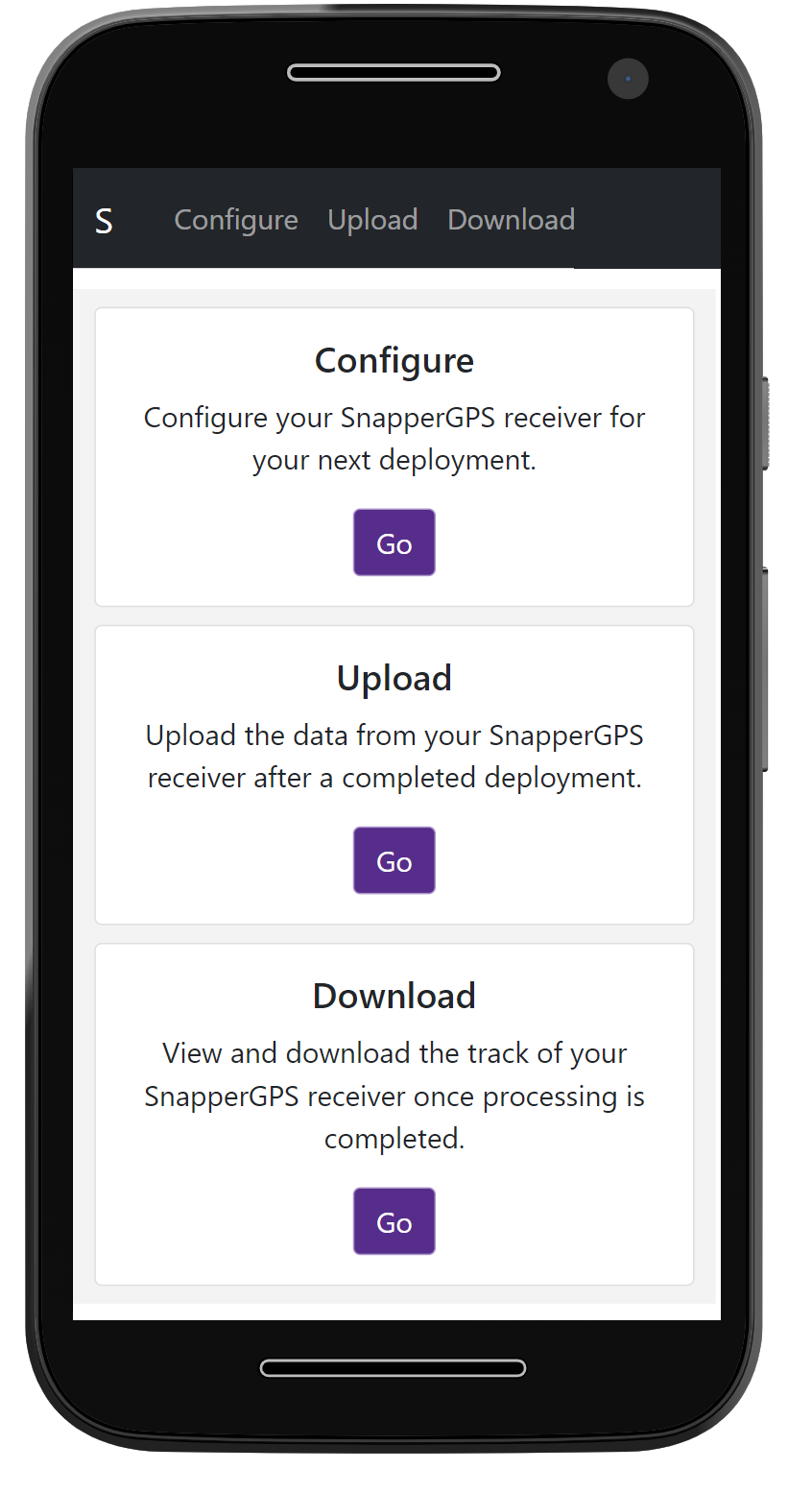}};
\node[block, right of=app] (snapshot-db)
    {\includegraphics[width=2em]{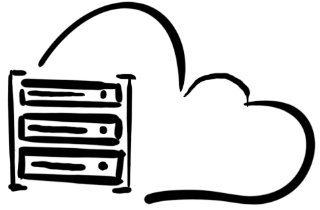}};
\node[close-block, right of=snapshot-db] (processing)
    {\includegraphics[width=2em]{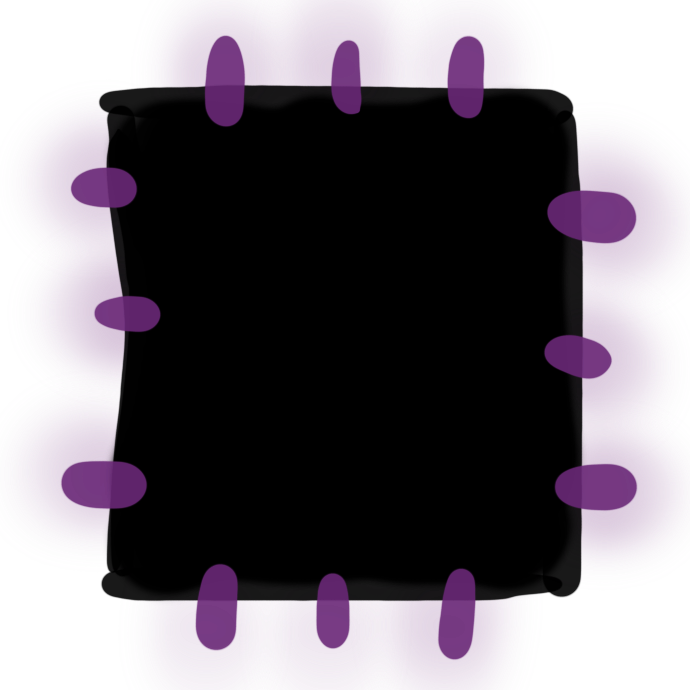}};
\node[close-block, right of=processing] (satellite-db)
    {\includegraphics[width=2em]{images/system_diagram/cloud_small_bw.png}};
\node[block, right of=satellite-db, node distance=7em] (external-db)
    {\includegraphics[width=2em]{images/system_diagram/cloud_small_bw.png}};
\node[block, below of=app, node distance=5.5em] (user)
    {\includegraphics[width=2em]{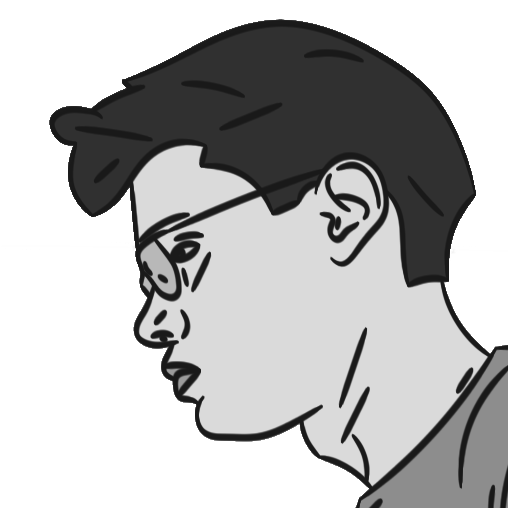}};
\draw [<-] ([yshift=+0.4em]receiver.east) -- node [arrow-txt, above] {configuration} ([yshift=+0.4em]app.west);
\draw [->] ([yshift=-0.4em]receiver.east) -- node [arrow-txt, below] {snapshots} ([yshift=-0.4em]app.west);
\draw [<->] (app) -- (user);
\draw [->] ([yshift=+0.4em]app.east) -- node [arrow-txt, above, xshift=-0.3em] {snapshots} ([yshift=+0.4em]snapshot-db.west);
\draw [<-] ([yshift=-0.4em]app.east) -- node [arrow-txt, below, xshift=-0.3em] {track} ([yshift=-0.4em]snapshot-db.west);
\draw [->] ([yshift=+0.4em]snapshot-db.east) -- ([yshift=+0.4em]processing.west);
\draw [<-] ([yshift=-0.4em]snapshot-db.east) -- ([yshift=-0.4em]processing.west);
\draw [<-] (processing) -- (satellite-db);
\draw [<-] (satellite-db) -- node [arrow-txt, below, align=center, xshift=+0.4em] {satellite\\data} (external-db);
\draw [->] (processing) -- ++ (0,-1.1) -- (3.3,-1.1) -/ node [arrow-txt, below, xshift=+2.1em, yshift=-0.4em] {notification} ([xshift=+1em]app.south);
\begin{pgfonlayer}{background}
    \path (snapshot-db.west |- processing.north)+(-0.2,3.2em) node (a) {};
    \path (processing.south -| satellite-db.east)+(+0.2,-2.8em) node (b) {};
    \path[fill=gray!5,rounded corners, draw=black!50, dashed]
        (a) rectangle (b);
\end{pgfonlayer}
\node[txt, above of=receiver] (receiver-txt)
    {\textbf{receiver}};
\node[txt, above of=app] (app-txt)
    {\textbf{app}};
\node[txt, above of=snapshot-db] (snapshot-db-txt)
    {snapshot\\\& track\\database};
\node[txt, above of=processing] (processing-txt)
    {snapshot\\processing};
\node[txt, above of=satellite-db] (satellite-db-txt)
    {satellite\\database};
\node[txt, above of=external-db] (external-db-txt)
    {3rd party\\database};
\node[txt, below of=user, node distance=2em] (user-txt)
    {user};
\node[txt, below of=processing, node distance=5.2em] (processing-txt)
    {\textbf{cloud-processing back-end}};
\end{tikzpicture}
\caption{The \snapper{} system comprising receiver, web application, and cloud-processing back-end.}
\label{fig:system}
\end{figure}

The actual processing of the GNSS signals snapshots, i.e., the location estimation, is done in the cloud rather than on the device itself (see Figure~\ref{fig:system}).
For this, we wrap the open-source snapshot GNSS algorithms developed by \cite{beuchert2021b} into an open-source Python back-end that runs on a server.
The back-end pulls snapshots that have been uploaded from a \snapper{} receiver from a relational database, processes them, and puts back location estimates.

\subsubsection{Web application}\label{sec:overview:implementation:web_application}

A front-end written in JavaScript functions as an interface between \snapper{} receiver, user, and back-end.
It is a progressive web app (PWA), which means that it can run in a browser window, but also as native app.
Additionally, PWAs load quickly, require little memory, are secure to use, and a service worker automatically keeps the app up-to-date.

The \snapper{} app communicates with a \snapper{} board via \mbox{WebUSB}, thus not requiring any driver installation.

A configuration view allows to read out various information, including
the battery voltage
and the number of signal snapshots on the receiver.
Furthermore, it is used to set parameters such as the device clock/time, the time interval between two consecutive snapshots, and the start and end date and time of the next deployment.
It can also be used to update the firmware via USB (see the firmware description presented earlier).

After a deployment, an upload view transfers snapshots and timestamps from the device to a server.

Finally, a download view shows the track on an interactive map once the back-end has calculated it and the user can download it in four different file formats.

\begin{figure}[tb]
\centerline{\includegraphics[width=1.0\columnwidth]{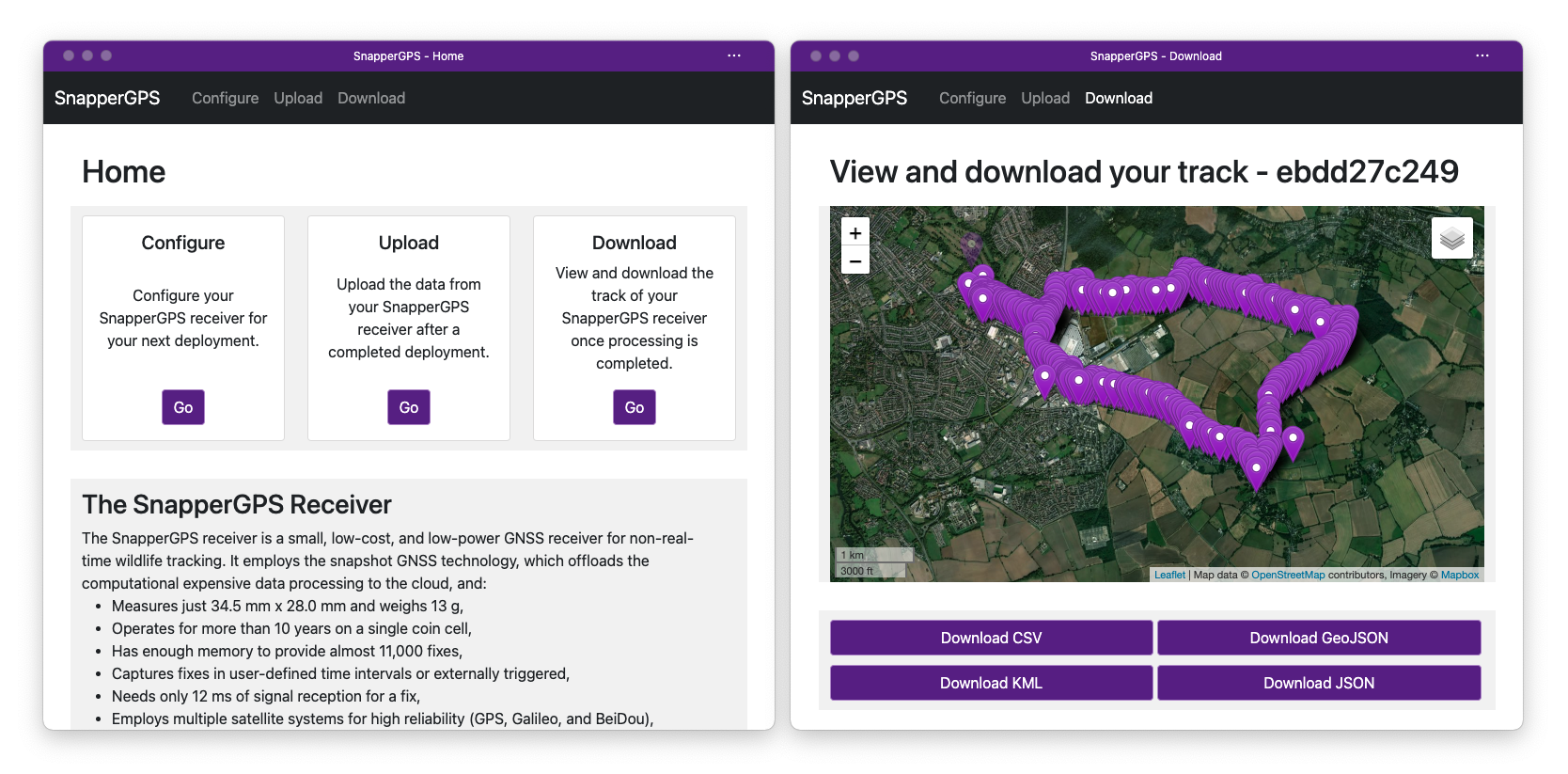}}
\caption{Example views of the \snapper{} app (\url{https://snappergps.info}).}
\label{fig:web}
\end{figure}

\section{(2) Quality Control}\label{sec:quality_control}

\subsection{Safety}\label{sec:quality_control:safety}

A lithium-ion polymer battery can overheat or catch fire when short-circuited, charged too quickly, overcharged, or damaged.
Always use a battery with in-built circuit protection and
follow the safety instructions of the supplier.%

\subsection{Calibration}\label{sec:quality_control:calibration}

A \snapper{} receiver does not require calibration despite the fact that it is made of low-cost components with relatively wide tolerances.
This is because the software automatically estimates and corrects errors.

The main components of concern are the two oscillators.
The \snapper{} receiver uses the 32.768~kHz oscillator to
time stamp snapshots.
The snapshot GNSS method requires timestamps with an accuracy of at least 1~min \citep{diggelen2009}.
However, the 32.768~kHz oscillators that we use have a frequency tolerance of at least $\pm$10~ppm, which can cause an error of over 5~min during a year-long deployment.
Our snapshot GNSS back-end mitigates this by estimating the time error for every single snapshot and using the previous estimate as initialisation for the next one.
Since the on-board time is synchronised with the browser time
during configuration, this propagation is sufficient to keep track of the time error during the location estimation process.

The synthesizer of the radio turns the 16.368 MHz reference frequency of the external TCXO into a frequency close to the GPS L1 frequency of 1575.42~MHz.
Therefore, frequency errors of the TCXO are amplified by almost a factor~100 and a frequency tolerance of $\pm$500~ppb can turn into a frequency offset of the front-end of more than 800~Hz, which shifts the intermediate frequency of the captured snapshots by the same amount.
However, for successful satellite acquisition, we need an accuracy that is a magnitude better than that, especially for Galileo and BeiDou satellites.
To avoid searching a large frequency space for the satellite signals, the back-end automatically estimates the offset of the receiver front-end for each uploaded dataset using a couple of snapshots from the beginning of the deployment \citep{beuchert2021a}.
Afterwards, it corrects the offset estimate over time based on the recorded temperatures and a linear model that relates temperature and frequency offset and which we fitted to a training dataset.

Both steps together allow us to obtain reliable localisation accuracy without manual calibration.

\subsection{General testing}\label{sec:quality_control:general_testing}

\subsubsection{Localisation accuracy}

\begin{figure}[tb]
  \raggedright
  \begin{minipage}[b]{0.48\linewidth}
    \centering
    \includegraphics[width=1.0\linewidth]{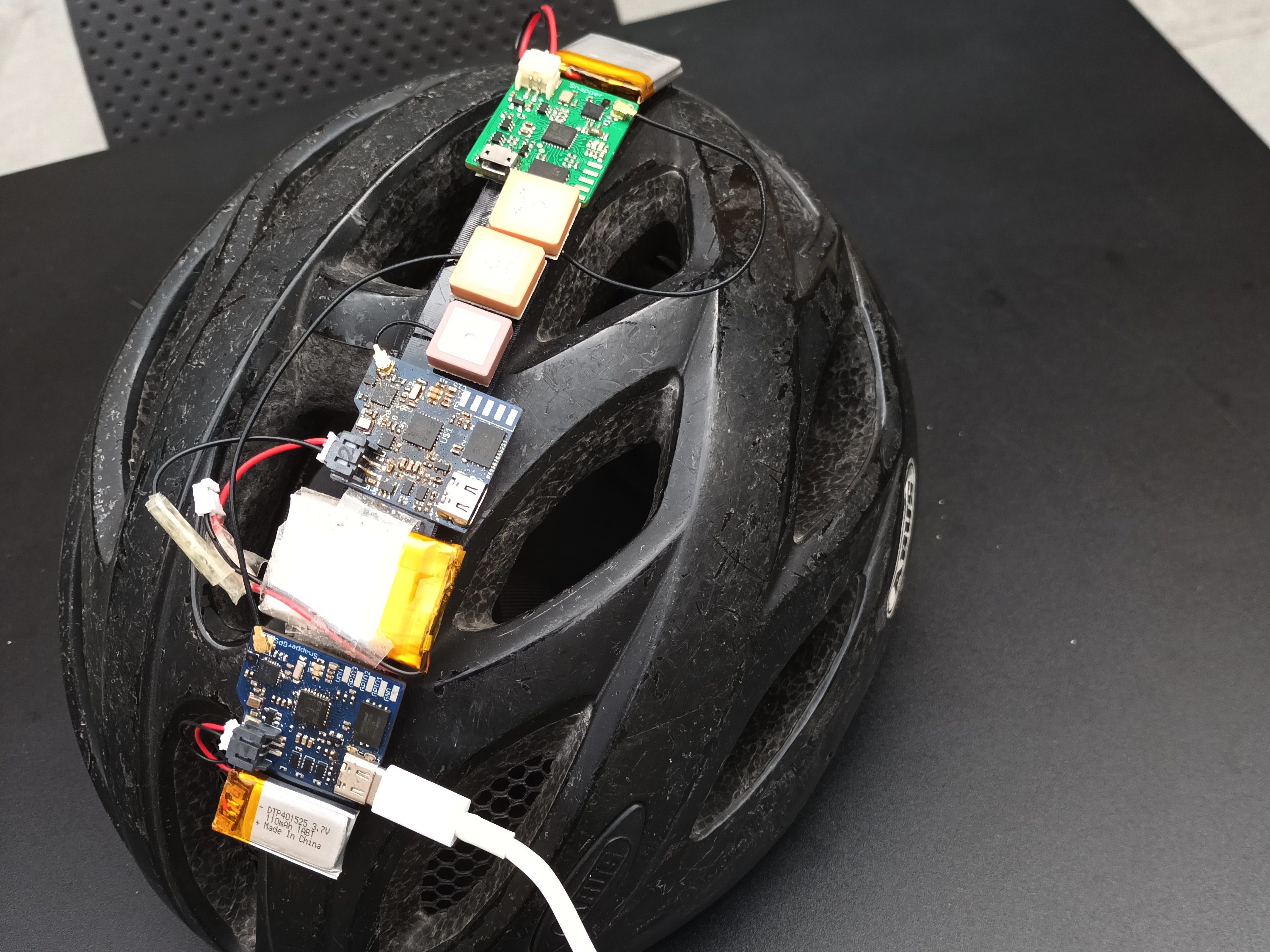} 
    \vspace{0.01\linewidth}
  \end{minipage}%
  \hspace{0.02\linewidth}
  \begin{minipage}[b]{0.48\linewidth}
    \centering
    \begin{overpic}[width=1.0\linewidth]{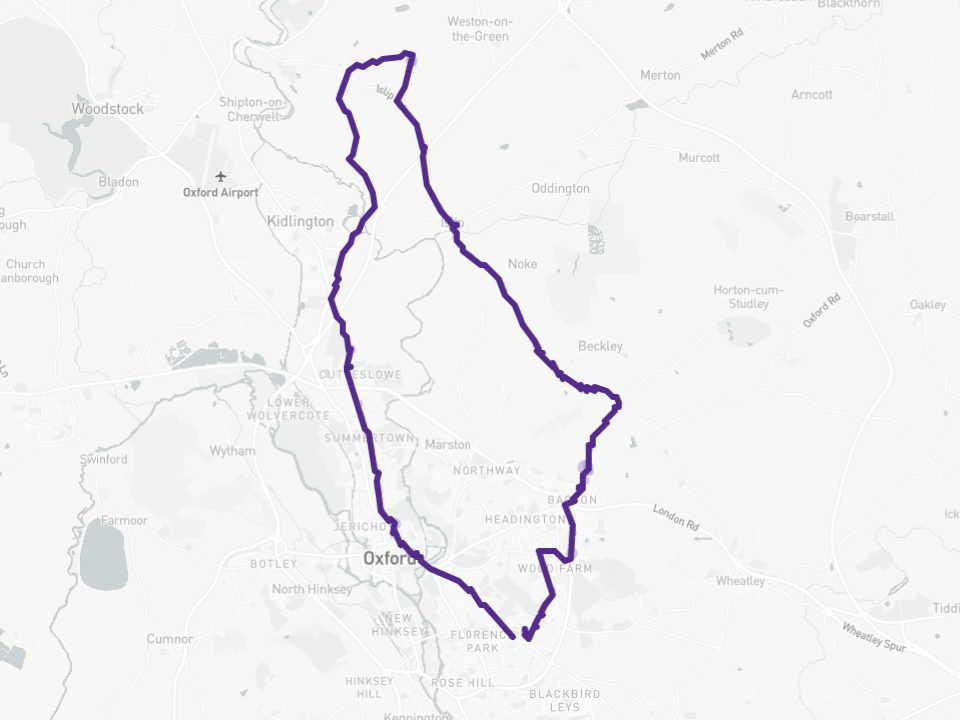}
        \put(2,2){\color{gray} \footnotesize Map data \copyright{} OpenStreetMap contributors}
        \put(2,8){\color{gray} \footnotesize Imagery \copyright{} Mapbox}
    \end{overpic}
    \vspace{0.01\linewidth}
  \end{minipage}
  \begin{minipage}[b]{0.48\linewidth}
    \centering
    \includegraphics[width=1.0\linewidth]{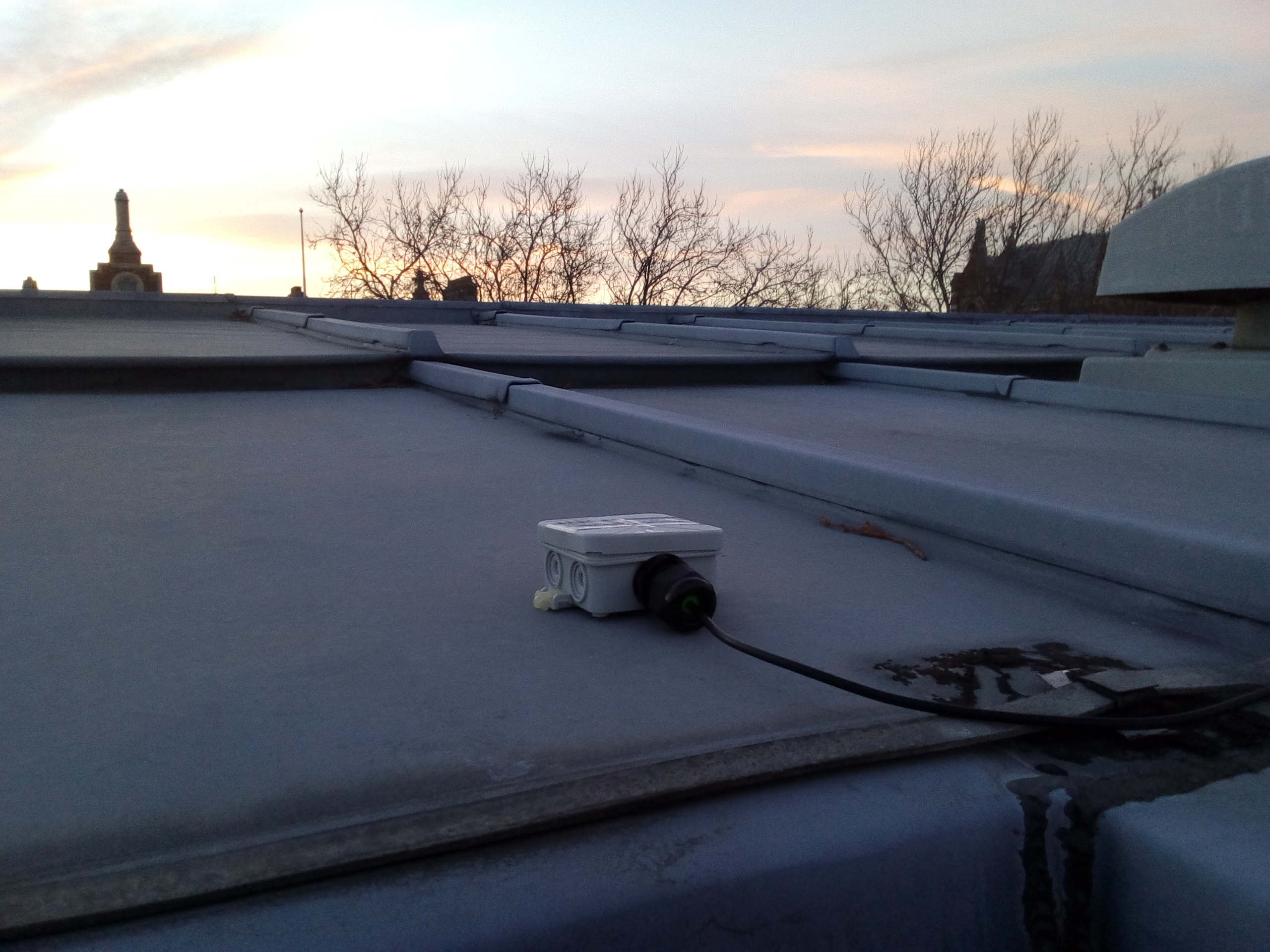} 
  \end{minipage}%
  \hspace{0.02\linewidth}
  \begin{minipage}[b]{0.48\linewidth}
    \centering
    \begin{overpic}[width=1.0\linewidth]{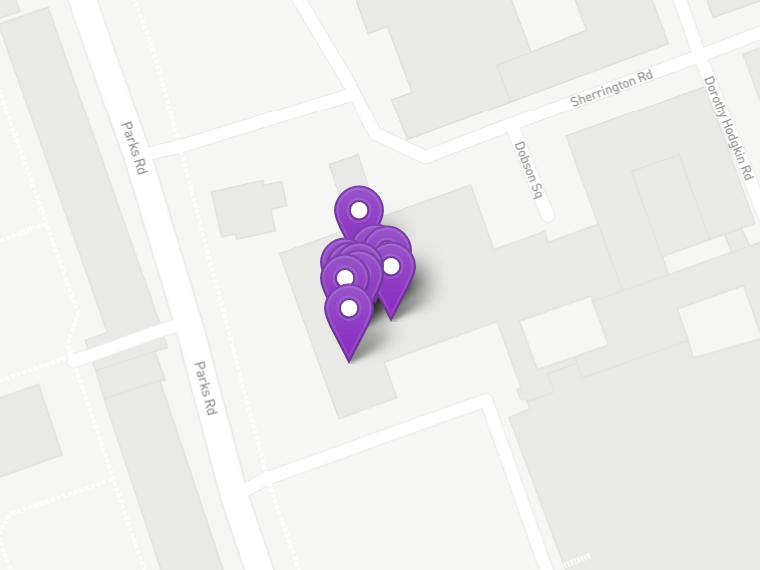}
        \put(2,2){\color{gray} \footnotesize Map data \copyright{} OpenStreetMap contributors}
        \put(2,8){\color{gray} \footnotesize Imagery \copyright{} Mapbox}
    \end{overpic}
  \end{minipage} 
\caption{Three \snapper{} receivers on a cycling helmet (top left), example track captured while cycling and walking in and around Oxford, UK (top right, from \url{https://snappergps.info/view?uploadid=8821aa36c3}), roof-mounted \snapper{} receiver for stationary data collection over six months (bottom left), and example set of ten fixes (bottom right).}
\label{fig:tests}
\end{figure}

To test the accuracy of the location estimates of a \snapper{} receiver, we advise users to just record some data outdoors at a location with known coordinates and then upload it via the \snapper{} app.
Then they can calculate the horizontal error between the estimated locations and the ground truth.
For comparison, we obtained median tracking errors of 11--12~m on an open test data collection that we established by ourselves using three \snapper{} receivers. It consists of four static and seven dynamic cycling tests in urban and rural environments with 3700~snapshots in total \citep{beuchert2021a, beuchert2021b}.
In addition, a receiver that we mounted on a roof top with good sky visibility collected about 8000~snapshots over more than five months and achieved a median error of 10~m (see Figure~\ref{fig:tests}).

\subsubsection{Power consumption}

Measuring the power/current consumption of a \snapper{} receiver is critical for long-term deployments to ensure that it will work for the desired operation time, although, none of our test boards had problems in this regard.
The current should be \SIrange[range-phrase=--]{1}{2}{\micro \ampere} if the board is sleeping and the maximum current should be around \SI{25}{\milli \ampere} when capturing a snapshot.
The charge consumption for a single snapshot should be <\SI{0.3}{\micro \ampere \hour}.

\section{(3) Application}\label{sec:application}

\subsection{Use cases}\label{sec:application:use_cases}

\snapper{} could be used for any location data logging application, however, the snapshot GNSS approach has advantages over traditional GNSS that make it particularly well suited for wildlife tracking. Most notable is the significantly reduced power consumption which means that smaller batteries can be used. Minimising size and weight of a tag is crucial when tagging wildlife to avoid unnecessary stress to the animal. Furthermore, a snapshot receiver is capable of taking a snapshot at any time and only requires milliseconds of clear signal (as opposed to multiple seconds with a traditional receiver). These attributes make the snapshot approach particularly well suited for marine settings, where animals may only surface sparsely and briefly.

\subsubsection{Marine deployment}

\begin{figure}[tb]
    \begin{minipage}[b]{0.48\linewidth}
        \includegraphics[width=\linewidth]{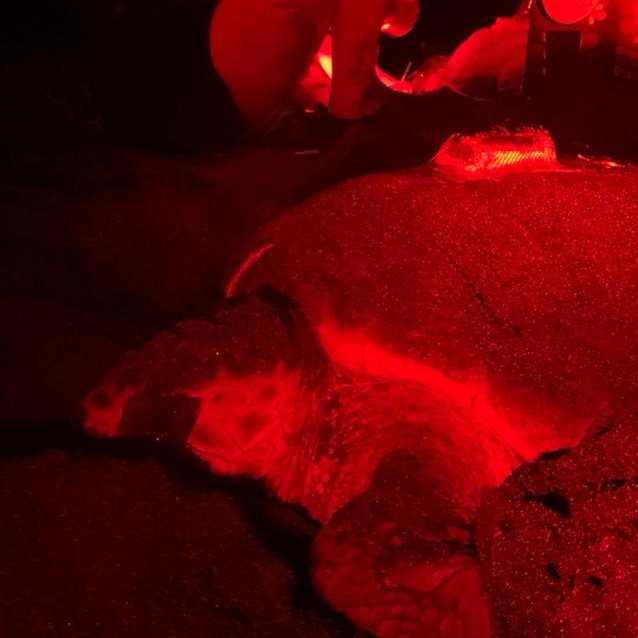} 
    \end{minipage}%
    \hspace{0.02\linewidth}
    \begin{minipage}[b]{0.48\linewidth}
        \begin{overpic}[width=\textwidth]{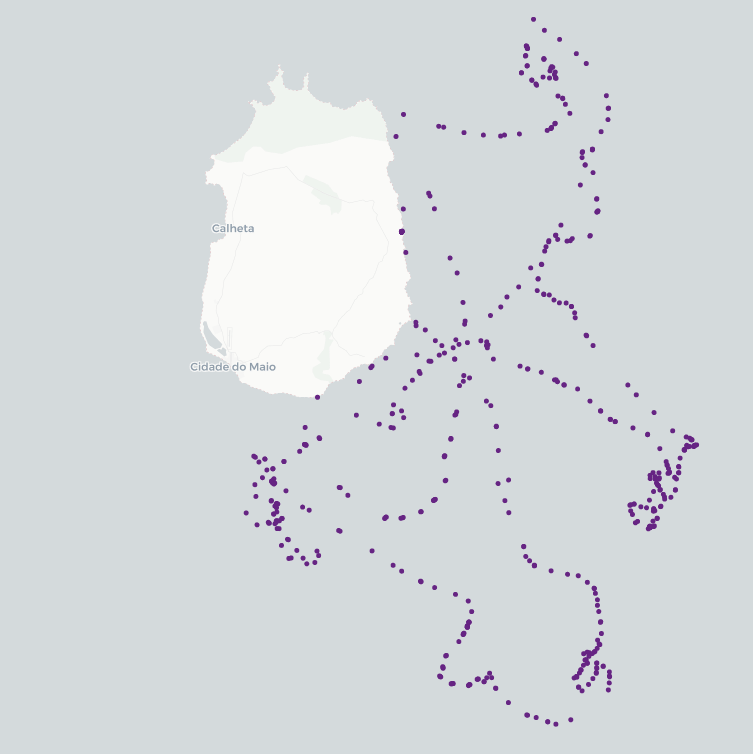}
        \put(2,2){\color{gray} \footnotesize Map data \copyright{} OpenStreetMap}
        \put(2,8){\color{gray} \footnotesize Imagery \copyright{} Mapbox}
        \end{overpic}
    \end{minipage}
    \caption{A \snapper{} tag on the back of a loggerhead sea turtle and a track captured over two weeks.}
    \label{fig:turtle}
\end{figure}

In summer 2021, \snapper{} tags were deployed on loggerhead sea turtles (Caretta caretta) on the island of Maio, Cape Verde~\citep{matthes2022snappergps}.
The receivers were placed in enclosures milled out of plastic and aluminium, which we tested to be waterproof to at least \SI{100}{\meter}. (We open-sourced an updated version of the enclosure design.)

The team tagged 20~females with \snapper{} tags while they were nesting on the beach.
Nine tags were recovered when the individuals returned to lay another clutch of eggs.
Whenever a snapshot was taken above the surface, \snapper{} reliably calculated a position.
GNSS signals are easily stopped by water and so any snapshots taken underwater do not contain any usable satellite information.
However, the \snapper{} signal processing chain can filter out these poor fixes, see Figure~\ref{fig:turtle}.

The obtained tracks are potentially helpful for identifying important habitats that may benefit from further protection.

\subsubsection{Avian deployment}
\begin{figure}[tb]
    \begin{minipage}[b]{0.48\linewidth}
        \includegraphics[width=\linewidth]{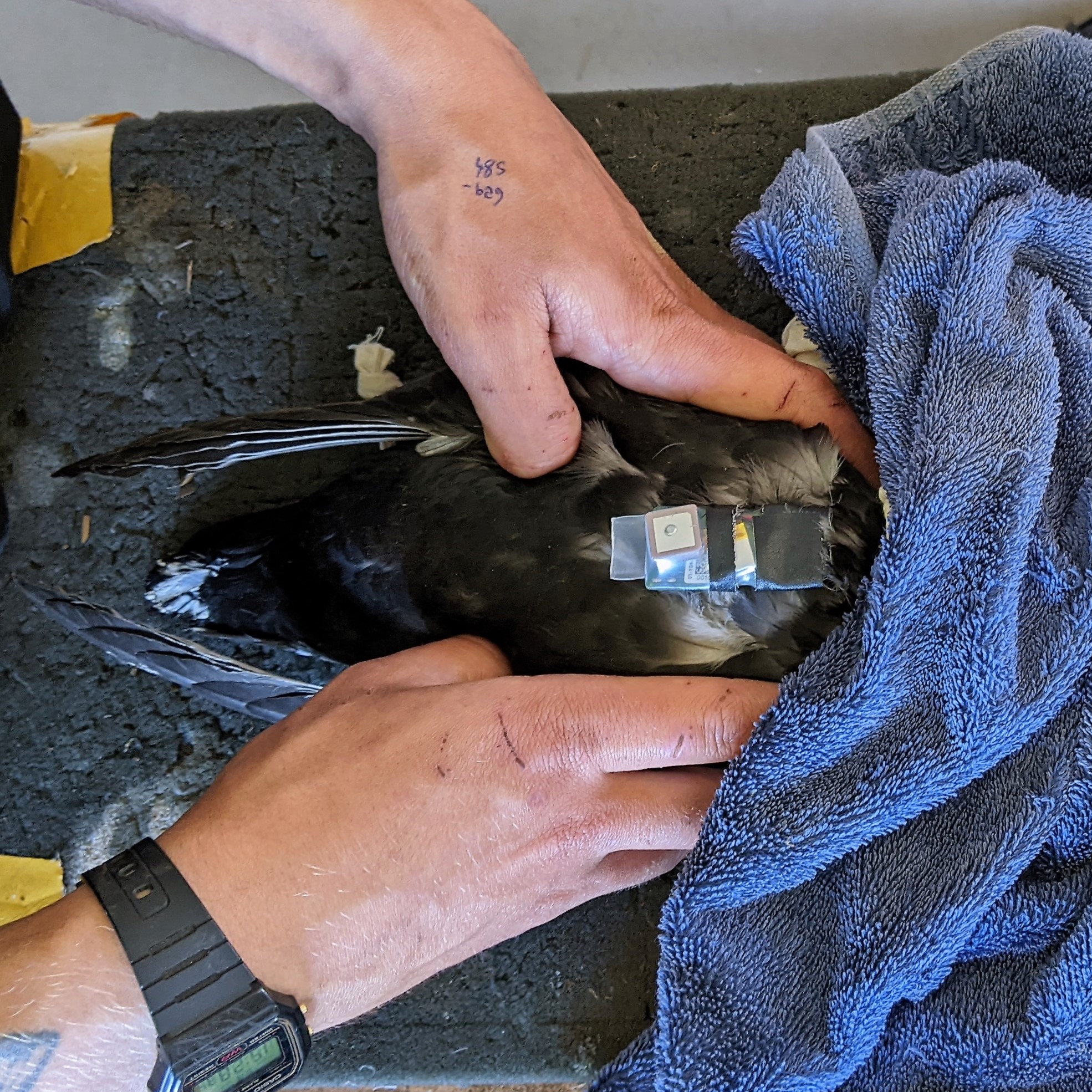} 
    \end{minipage}%
    \hspace{0.02\linewidth}
    \begin{minipage}[b]{0.48\linewidth}
        \begin{overpic}[width=\textwidth]{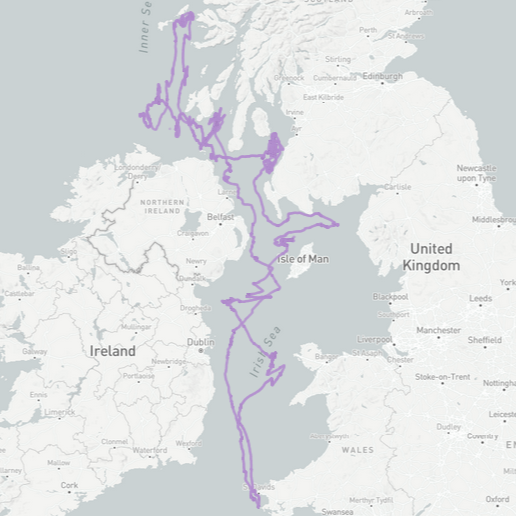}
        \put(2,2){\color{gray} \footnotesize Map data \copyright{} OpenStreetMap}
        \put(2,8){\color{gray} \footnotesize Imagery \copyright{} Mapbox}
        \end{overpic}
    \end{minipage}
    \caption{A custom \snapper{} tag on the back of a Manx shearwater and a recorded track.}
    \label{fig:bird}
\end{figure}

In May 2022, a custom variant of \snapper{} was deployed on Manx shearwaters (Puffinus puffinus) nesting on Skomer, Wales and the Copeland Islands, Northern Ireland.
These birds weigh between \SI{400}{\gram} and \SI{450}{\gram} \citep{brooke2013manx}, so minimising the weight of the tag is particularly important. To achieve this, we reduced the size of the PCB by removing the GPIO pads and the JST connector for the lithium-ion polymer battery. The battery is soldered on instead. To avoid having to remove the battery for charging, we added a charging circuit to the PCB. This allows the tag to be charged using the USB port. The altered tag with a \SI{40}{\milli \ampere \hour} lithium-ion polymer battery and an APAM1368YB13V3.0 antenna from Abracon LLC weighs \SI{8.3}{\gram}. %

The receivers were waterproofed using heat-shrink tubing.
The tubing was shrunk with a heat gun and then the ends were sealed with flat needle-nose pliers while the plastic was still warm.
In total, 21~tags were deployed and 19 were recovered over the incubation and chick rearing period.
The recorded data has high enough temporal resolution to identify behaviour.
In particular, foraging behaviour shows up in the data as erratic movement which clearly distinguishes it from sustained flight or rafting on the surface.
This makes it possible to monitor behaviour and locate major foraging sites of the species during the nesting season without having to co-deploy other sensors such as accelerometers. 

See Figure~\ref{fig:bird} for a photograph of a \snapper{} receiver on a bird and a plot of one recorded track. %

\subsection{Reuse potential and adaptability}

SnapperGPS, as it is presented here, is designed to be general-purpose and low-cost. However, as demonstrated in the previous section, the SnapperGPS hardware can be customised for specific applications. This is a key advantage of a fully open-source solution. %

The open-source design has multiple GPIO pins to extend functionality of the tag without altering the PCB layout. These can be used to connect a sensor such as an accelerometer, a calibrated thermometer or a light sensor. The additional sensor data can then be logged alongside the GNSS data. Such additional sensors can also be used to trigger snapshots.

Other potentially useful changes to the PCB design may include switching out the flash storage for a larger alternative (e.g. a microSD card), and swapping the power source from a LiPo battery to something that is more readily available (e.g. a coin cell).

\section{(4) Build Details}\label{sec:build_details}

\subsection{Availability of materials and methods}\label{sec:build_details:materials_and_methods}

Table~\ref{tab:bill} presents the bill of materials (BOM) for the electronic components of a \snapper{} receiver.
The price of \$21 for a single device in a batch of 100 is significantly cheaper than both, existing commercial and open-source GNSS receivers for wildlife tracking, which usually cost a few hundred USD.

\begin{table}[!htbp]
    \centering
    \caption{Bill of materials for one, 100, and 1000 \snapper{} receivers.}
    \label{tab:bill}
    \resizebox{\columnwidth}{!}{%
        \begin{tabular}{l r l r r r r}
            \hline
            \multirow{ 2}{*}{\textbf{Component}}& \multirow{ 2}{*}{\textbf{Value}}& \multirow{ 2}{*}{\textbf{Part}}& \multirow{ 2}{*}{\textbf{Qty}} & \multicolumn{3}{c}{\textbf{Price [USD]}}  \\
            &&&& \textbf{one} & \textbf{100} & \textbf{1000} \\
            \hline
            Ceramic capacitor & 100 nF & GRM155R60J104KA01D & 10 & 0.33 & 14.60 & 81.70 \\
            Ceramic capacitor & 10 nF & GRM155R60J103KA01D & 5 & 0.50 & 4.43 & 29.10 \\
            Ceramic capacitor & 100 pF & GRM1555C1H101JA01J & 1 & 0.10 & 0.81 & 4.54 \\
            Ceramic capacitor & 22 pF & GRM1555C1H220JA01D & 1 & 0.10 & 0.81 & 4.57 \\
            Ceramic capacitor & 18 pF & GRM1555C1H180JA01D & 2 & 0.20 & 2.32 & 13.00 \\
            Multi-layer ceramic capacitor & 1 uF & GRM188R60J105KA01D & 4 & 0.40 & 11.00 & 61.84 \\
            Multi-layer ceramic capacitor & 4.7 uF & GRM185R60J475ME15D & 1 & 0.30 & 11.67 & 72.02 \\
            Multi-layer ceramic capacitor & 10 uF & GRM188R60J106KE47D & 1 & 0.15 & 5.07 & 29.44 \\
            Fixed inductor & 39 nH & 0402HS-390EKTS & 1 & 0.26 & 15.68 & 109.72 \\
            Ferrite bead & ~ & BLM15HB221SH1D & 1 & 0.10 & 4.36 & 25.08 \\
            Resistor & 15 \textohm & RC0402FR-0715RL & 2 & 0.20 & 1.32 & 5.98 \\
            Resistor & 1 k\textohm & RC0402FR-071KL & 2 & 0.20 & 1.32 & 5.98 \\
            Resistor & 4.7 k\textohm & RC0402FR-074K7L & 3 & 0.30 & 1.98 & 7.77 \\
            Resistor & 5.1 k\textohm & RC0402FR-075K1L & 2 & 0.20 & 1.32 & 5.98 \\
            Resistor & 6.8 k\textohm & RC0402FR-076K8L & 1 & 0.10 & 0.66 & 2.99 \\
            Resistor & 10 k\textohm & RC0402FR-0710KL & 2 & 0.20 & 1.32 & 5.98 \\
            Resistor & 100 k\textohm & RC0402FR-07100KL & 7 & 0.70 & 4.62 & 14.98 \\
            Crystal & 32.768 kHz & Q 0,032768-JTX310-12,5-10-T1-HMR-50K-LF & 1 & 0.86 & 62.70 & 495.00 \\
            Yellow-green LED & 572 nm & SML-D12M1WT86 & 1 & 0.22 & 5.51 & 39.86 \\
            Red LED & 620 nm & SML-D12U1WT86 & 1 & 0.22 & 5.51 & 39.86 \\
            Temp. compensated oscillator & 16.368 MHz & D32G-016.368M & 1 & 8.00 & 700.00 & 7000.00 \\
            GPS receiver & ~ & SE4150L-R & 1 & 0.86 & 86.40 & 864.00 \\
            SAW filter & 1.57542 GHz & AFS20A42-1575.42-T3 & 1 & 1.05 & 71.37 & 543.74 \\
            Microcontroller & ~ & EFM32HG310F64G-C-QFN32R & 1 & 5.91 & 435.04 & 3703.46 \\
            TVS diode & 5.5 V & PRTR5V0U2AX,235 & 1 & 0.66 & 42.47 & 257.87 \\
            NAND flash memory & 512 Mbit & W25N512GVEIG & 1 & 2.90 & 228.68 & 2051.92 \\
            U.FL connector jack & 50 \textohm & 1909763-1 & 1 & 0.51 & 36.55 & 595.90 \\
            Mosfet array N/P-channel & ~ & TT8M1TR & 3 & 1.56 & 86.58 & 487.08 \\
            Linear voltage regulator & ~ & S-1318D33-M5T1U4 & 1 & 1.58 & 109.94 & 908.20 \\
            USB-C receptacle connector & ~ & DX07S016JA1R1500 & 1 & 1.67 & 129.65 & 1080.44 \\
            JST connector header & ~ & S2B-PH-SM4-TB & 1 & 0.54 & 39.08 & 339.84 \\
            \hline
            \multicolumn{3}{c}{\textbf{Total}} & & 31.55 & 2124.54 & 18900.69 \\
            \hline
        \end{tabular}
    }
\end{table}

\subsection{Ease of build}\label{sec:build_details:ease_of_build}

We intend to support making the \snapper{} hardware available through group purchasing websites.
This will be more cost- and time-effective for most potential users.

\subsection{Operating software and peripherals}\label{sec:build_details:software}

See sections \nameref{sec:overview:implementation:firmware} and \nameref{sec:overview:implementation:web_application}.

The \snapper{} app works in the \textsc{Microsoft Edge} and \textsc{Google Chrome} browsers.
Optionally, the user can also install it as a native app,
either through the website \url{https://snappergps.info}, the \textsc{Microsoft Store} (\url{https://apps.microsoft.com/store/detail/snappergps/9P9RPRS6LSMM}), or \textsc{Google Play} (\url{https://play.google.com/store/apps/details?id=com.herokuapp.snapper_gps.twa}).

\subsection{Dependencies}\label{sec:build_details:dependencies}

The \snapper{} hardware has no dependencies.
We provide all firmware dependencies together with the firmware.
We list all dependencies of the web app in the respective repositories and they can be installed via pip (Python) or npm (JavaScript).
No dependencies are closed-source and all are free software.

\subsection{Hardware documentation and files location}\label{sec:build_details:hardware_documentation_and_files}
Name: snappergps-pcb \\  %
Persistent identifier: \url{https://github.com/SnapperGPS/snappergps-pcb} \\  %
Licence: Solderpad Hardware License v2.1\\
Publishers: Jonas Beuchert, Amanda Matthes, Alex Rogers\\
Date published: 10/06/22

\underline{Software code repositories}  %
\\  %
Names: snappergps-firmware, snappergps-app, and snappergps-backend \\  %
Identifiers: \url{https://github.com/SnapperGPS/snappergps-firmware}, \url{https://github.com/SnapperGPS/snappergps-app}, and \url{https://github.com/SnapperGPS/snappergps-backend} \\  %
Licence: MIT \\  %
Date published: 10/06/22

\section{(5) Discussion}\label{sec:discussion}

\subsection{Conclusions}\label{sec:discussion:conclusions}

This paper presents \snapper{}, a location data logging system designed for wildlife tracking. It consists of three parts: (i) a GNSS receiver, (ii) a web application that is used to configure it, and (iii) a cloud-based data processing platform. The receiver is small, low-cost and energy-efficient enough to run on small batteries for multiple years. \snapper{} has been tested in multiple controlled static and dynamic tests in urban and rural environments. Additionally, \snapper{} has already been used for tracking sea turtles and sea birds in the wild. These deployments have demonstrated that \snapper{} is a viable wildlife tracking system.

\subsection{Future Work}\label{sec:discussion:future_work}

Future work could include: altering the PCB design to further reduce the size and weight of the receiver; switching to a power source that is more readily available; increasing the on-board storage to allow for more snapshots per deployment; and adding commonly used sensors to log additional data (e.g. sound, light, acceleration).

\section{Paper author contributions}\label{sec:contributions}%

JB: Firmware development, web application development, hardware design, prototype assembly, testing, deployments, documentation, paper writing\\
AM: Prototype assembly, hardware design, testing, deployments, deployment organisation, documentation, paper writing\\
AR: Concept, hardware design, prototype assembly, firmware development, web application development, testing, project supervision

\section{Acknowledgements}\label{sec:acknowledgements}%

The development of \snapper{} was supported by Andy Hill and Peter Prince from Open Acoustic Devices. The \snapper{} deployment on loggerhead sea turtles was conducted in cooperation with the Arribada Initiative and the Maio Biodiversity Foundation. Special thanks to Alasdair Davies, Juan Patino-Martinez and Rocio Moreno. We also thank Joe Morford, Patrick Lewin, Katrina Davies, Lewis Fisher-Reeves and Tim Guilford of the Oxford Navigation Group, which conducted the \snapper{} deployment on Manx Shearwaters.

\section{Funding statement}\label{sec:funding_statement} %

SnapperGPS was funded by an EPSRC IAA Technology Fund (D4D00010-BL14). Jonas Beuchert and Amanda Matthes are funded by the EPSRC Centre for Doctoral Training in Autonomous Intelligent Machines and Systems
(DFT00350-DF03.01, DFT00350-DF03.05).

\section{Competing interests}\label{sec:competing_interests} %

The authors declare that they have no competing interests.

\bibliography{paper.bib}

\end{document}